\documentclass{article}
\usepackage{nips06,times}
\usepackage{latexsym}
\usepackage[dvips]{graphicx} 
\def\alt{\mathrel{\hbox{\rlap{\hbox{\lower4pt\hbox{$\sim$}}}
\hbox{\raisebox{0.4ex}{\hspace*{-0.05in}$<$}}}}}

\def\agt{\mathrel{\hbox{\rlap{\hbox{\lower4pt\hbox{$\sim$}}}
\hbox{\raisebox{0.4ex}{\hspace*{-0.05in}$>$}}}}}
\title{Comparison of objective functions for estimating
  linear-nonlinear models}

\author{
Tatyana O. Sharpee\\
Computational Neurobiology Laboratory, \\the Salk Institute for Biological Studies, La Jolla, CA 92037 \\
\texttt{sharpee@salk.edu} 
}

\begin{document}

\maketitle

\begin{abstract}
  This paper compares a family of methods for characterizing neural
  feature selectivity with natural stimuli in the framework of the
  linear-nonlinear model. In this model, the neural firing rate is a
  nonlinear function of a small number of relevant stimulus
  components. The relevant stimulus dimensions can be found by
  maximizing one of the family of objective functions, R\'enyi
  divergences of different orders~\cite{Paninski03,Sharpee04}. We show
  that maximizing one of them, R\'enyi divergence of order 2, is
  equivalent to least-square fitting of the linear-nonlinear model to
  neural data. Next, we derive reconstruction errors in relevant
  dimensions found by maximizing R\'enyi divergences of arbitrary
  order in the asymptotic limit of large spike numbers. We find that
  the smallest errors are obtained with R\'enyi divergence of order 1,
  also known as Kullback-Leibler divergence. This corresponds to
  finding relevant dimensions by maximizing mutual
  information~\cite{Sharpee04}. We numerically test how these
  optimization schemes perform in the regime of low signal-to-noise
  ratio (small number of spikes and increasing neural noise) for model
  visual neurons. We find that optimization schemes based on either
  least square fitting or information maximization perform well even
  when number of spikes is small. Information maximization provides
  slightly, but significantly, better reconstructions than least
  square fitting. This makes the problem of finding relevant
  dimensions, together with the problem of lossy
  compression~\cite{Tishby2007}, one of examples where
  information-theoretic measures are no more data limited than those
  derived from least squares.

\end{abstract}

\section{Introduction}
\label{sec:intro}

The application of system identification techniques to the study of
sensory neural systems has a long history. One family of approaches
employs the dimensionality reduction idea: while inputs are typically
very high-dimensional, not all dimensions are equally important for
eliciting a neural
response~\cite{deBoer,Hunter_Korenberg_1986b,stcov,Marmarelis_1997,rob+bill-feature}.
The aim is then to find a small set of dimensions $\{ \hat e_1, \,
\hat e_2,\, \ldots\} $ in the stimulus space that are relevant for
neural response, without imposing, however, a particular functional
dependence between the neural response and the stimulus components $\{
s_1,\, s_2,\, \ldots \}$ along the relevant dimensions:
\begin{equation}
\label{ior}
P({\rm spike}|{\bf s})=P({\rm spike}) g(s_1,s_2, ...,s_K), 
\end{equation}

If the inputs are Gaussian, the last requirement is not important,
because relevant dimensions can be found without knowing a correct
functional form for the nonlinear function $g$ in
Eq.~(\ref{ior}). However, for non-Gaussian inputs a wrong assumption
for the form of the nonlinearity $g$ will lead to systematic errors in
the estimate of the relevant dimensions
themselves~\cite{Ringach97,Hunter_Korenberg_1986b,Paninski03,Sharpee04}. The
larger the deviations of the stimulus distribution from a Gaussian,
the larger will be the effect of errors in the presumed form of the
nonlinearity function $g$ on estimating the relevant
dimensions. Because inputs derived from a natural environment, either
visual or auditory, have been shown to be strongly non-Gaussian~\cite{Ruderman}, we
will concentrate here on system identification methods suitable for
either Gaussian or non-Gaussian stimuli.

To find the relevant dimensions for neural responses probed with
non-Gaussian inputs, Hunter and Korenberg proposed an iterative
scheme~\cite{Hunter_Korenberg_1986b} where the relevant dimensions are
first found by assuming that the input--output function $g$ is linear.
Its functional form is then updated given the current estimate of the
relevant dimensions. The inverse of $g$ is then used to improve the
estimate of the relevant dimensions. This procedure can be improved
not to rely on inverting the nonlinear function $g$ by formulating
optimization problem exclusively with respect to relevant
dimensions~\cite{Paninski03,Sharpee04}, where the nonlinear function
$g$ is taken into account in the objective function to be optimized. A
family of objective functions suitable for finding relevant dimensions
with natural stimuli have been proposed based on R\'enyi
divergences~\cite{Paninski03} between the the probability
distributions of stimulus components along the candidate relevant
dimensions computed with respect to all inputs and those associated
with spikes. Here we show that the optimization problem based on the
R\'enyi divergence of order 2 corresponds to least square fitting of
the linear-nonlinear model to neural spike trains. The
Kullback-Leibler divergence also belongs to this family and is the
R\'enyi divergence of order 1.  It quantifies the amount of mutual
information between the neural response and the stimulus components
along the relevant dimension~\cite{Sharpee04}.  The optimization
scheme based on information maximization has been previously proposed
and implemented on model~\cite{Sharpee04} and real
cells~\cite{Sharpee06}. Here we derive asymptotic errors for
optimization strategies based on R\'enyi divergences of arbitrary
order, and show that relevant dimensions found by maximizing
Kullback-Leibler divergence have the smallest errors in the limit of
large spike numbers compared to maximizing other R\'enyi divergences,
including the one which implements least squares. We then show in
numerical simulations on model cells that this trend persists even for
very low spike numbers.

\section{Variance as an Objective Function} 
\label{sub:varmin}
One way of selecting a low-dimensional model of neural response is to
minimize a $\chi^2$-difference between spike probabilities measured
and predicted by the model after averaging across all inputs ${\bf
s}$:
\begin{equation}
  \chi^2[{\bf v}]=\int d{\bf s} P({\bf s}) \left [ \frac{P({\rm
        spike}|{\bf s})}{P({\rm spike})}- \frac{P({\rm spike}|{\bf
        s}\cdot {\bf v})}{P({\rm spike})}\right]^2,
\end{equation}
where dimension ${\bf v}$ is the relevant dimension for a given model
described by Eq.~(\ref{ior}) [multiple dimensions could also be
used, see below]. Using the Bayes' rule and rearranging terms, we get:
\begin{equation}
\label{H}
\chi^2[{\bf v}]=\hspace*{-0.07in}\int\hspace*{-0.07in} d{\bf s} P({\bf s}) \left [ \frac{P({\bf s}|{\rm
      spike})}{P({\bf s})}- \frac{P({\bf s}\cdot {\bf v}|{\rm
      spike})}{P({\bf s}\cdot{\bf v})}\right]^2\hspace*{-0.07in}=\hspace*{-0.07in}\int\hspace*{-0.07in} d{\bf s} \frac{[P({\bf s}|{\rm
    spike})]^2}{P({\bf s})} - \int \hspace*{-0.07in} dx  \frac{[P_{\bf v}(x|{\rm
    spike})]^2}{P_{\bf v}(x)}. 
\end{equation}
In the last integral averaging has been carried out with respect to
all stimulus components except for those along the trial direction
${\bf v}$, so that integration variable $x={\bf s}\cdot {\bf v}$. 
Probability distributions $P_{\bf v}(x)$ and $P_{\bf v}(x|{\rm
  spike})$ represent the result of this averaging across all presented
stimuli and those that lead to a spike, respectively:
\begin{equation}
\label{Pv}
  P_{\bf v}(x)=\int d{\bf s}  P({\bf s}) \delta(x-{\bf s}\cdot {\bf
  v}), \quad   P_{\bf v}(x|{\rm spike})=\int d{\bf s}  P({\bf s}|{\rm spike}) \delta(x-{\bf
    s}\cdot {\bf v}),
\end{equation}
where $\delta(x)$ is a delta-function.  In practice, both of the
averages (\ref{Pv}) are calculated by bining the range of projections
values $x$ and computing histograms normalized to unity. Note that if
there multiple spikes are sometimes elicited, the probability
distribution $P(x|{\rm spike})$ can be constructed by weighting the
contribution from each stimulus according to the number of spikes it
elicited.

If neural spikes are indeed based on one relevant dimension, then this
dimension will explain all of the variance, leading to $\chi^2=0$. For
all other dimensions ${\bf v}$, $\chi^2[{\bf v}]>0$. Based on
Eq.~(\ref{H}), in order to minimize $\chi^2$ we need to maximize

\begin{equation}
\label{Fv}
F[{\bf v}]=\int dx P_{\bf v}(x) \left[\frac{ P_{\bf v}(x|{\rm spike})}{P_{\bf v}(x)}\right ]^2, 
\end{equation}
which is a R\'enyi divergence of order 2 between probability
distribution $P_{\bf v}(x|{\rm spike})$ and $P_{\bf v}(x)$, and are
part of a family of $f$-divergences measures that are based on a
convex function of the ratio of the two probability distributions
(instead of a power $\alpha$ in a R\'enyi divergence of order
$\alpha$)~\cite{Ali_Silvey_1966,Csiszar_1967,Paninski03}.  For
optimization strategy based on R\'enyi divergences of order $\alpha$,
the relevant dimensions are found by maximizing:
\begin{equation}
\label{Falpha}
F^{(\alpha)}[ {\bf v}]=\frac{1}{\alpha-1}\int dx P_{\bf v}(x) \left[\frac{ P_{\bf v}(x|{\rm
spike})}{P_{\bf v}(x)}\right ]^\alpha.
\end{equation}
 By comparison, when the relevant dimension(s) are found by maximizing
information~\cite{Sharpee04}, the goal is to maximize Kullback-Leibler
divergence, which can be obtained by taking a formal limit $\alpha \to
1$:
\begin{equation}
\label{Iv}
I[ {\bf v}]=\int dx P_{\bf v}(x)\frac{ P_{\bf v}(x|{\rm spike})
}{P_{\bf v}(x)} \ln \frac{ P_{\bf v}(x|{\rm spike}) }{P_{\bf
v}(x)}=\int dx P_{\bf v}(x|{\rm spike}) \ln \frac{ P_{\bf v}(x|{\rm
spike}) }{P_{\bf v}(x)}.
\end{equation}

Returning to the variance optimization, the maximal value for $F[{\bf
v}]$ that can be achieved by any dimension ${\bf v}$ is:
\begin{equation}
\label{fmax}
  F_{\rm max}=\int d{\bf s}\frac{\left[P({\bf
    s}|{\rm spike})\right]^2}{P({\bf s})}.
\end{equation}
It corresponds to the variance in the firing rate averaged across
different inputs (see Eq.~(\ref{exp_fmax}) below). Computation of the
mutual information carried by the individual spike about the stimulus
relies on similar integrals. Following the procedure outlined for
computing mutual information~\cite{B00a}, one can use the Bayes'
rule and the ergodic assumption to compute $F_{\rm max}$ as a
time-average:
\begin{equation}
\label{exp_fmax}
  F_{\rm max}=\frac{1}{T}\int dt \left[\frac{r(t)}{\bar r}\right]^2,
\end{equation}
where the firing rate $r(t)=P({\rm spike}|{\bf s})/\Delta t$ is
  measured in time bins of width $\Delta t$ using multiple repetitions
  of the same stimulus sequence . The stimulus ensemble should be
  diverse enough to justify the ergodic assumption [this could be
  checked by computing $F_{\rm max}$ for increasing fractions of the
  overall dataset size].  The average firing rate $\bar r=P({\rm
  spike})/\Delta t$ is obtained by averaging $r(t)$ in time.

The fact that $F[{\bf v}]<F_{\rm max}$ can be seen either by simply
noting that $\chi^2[{\bf v}]\geq 0$, or from the data processing
inequality, which applies not only to Kullback-Leibler divergence, but
also to R\'enyi
divergences~\cite{Ali_Silvey_1966,Csiszar_1967,Paninski03}. In other
words, the variance in the firing rate explained by a given dimension
$F[{\bf v}]$ cannot be greater than the overall variance in the firing
rate $F_{\rm max}$. This is because we have averaged over all of the
variations in the firing rate that correspond to inputs with the same
projection value on the dimension ${\bf v}$ and differ only in
projections onto other dimensions.

Optimization scheme based on R\'enyi divergences of different orders
have very similar structure. In particular, gradient could be
evaluated in a similar way:
\begin{equation}
\label{gradF}
\nabla_{\bf v} F^{(\alpha)} = \frac{\alpha}{\alpha-1}\int dx  P_{\bf v}(x|{\rm spike}) \left[\langle
{\bf s}|x, {\rm spike}\rangle-\langle {\bf s}|x\rangle \right] 
\frac{d}{dx}\left[ \left( \frac{P_{\bf v}(x|{\rm spike})}{P_{\bf v}(x)}\right)^{\alpha-1}\right],
\end{equation}
where $ \langle {\bf s}|x,{\rm spike}\rangle=\int d {\bf s}\, {\bf
  s}\delta(x-{\bf s}\cdot{\bf v})P({\bf s}|{\rm spike})/P(x|{\rm
  spike})$, and similarly for $\langle {\bf s}|x\rangle$. The gradient
  is thus given by a weighted sum of spike-triggered averages $\langle
  {\bf s}|x,{\rm spike}\rangle- \langle {\bf s}|x \rangle$ conditional
  upon projection values of stimuli onto the dimension ${\bf v}$
  for which the gradient of information is being evaluated. The
  similarity of the structure of both the objective functions and
  their gradients for different R\'enyi divergences means that numeric
  algorithms can be used for optimization of R\'enyi divergences of
  different orders. Examples of possible algorithms have been
  described~\cite{Paninski03,Sharpee04,Sharpee06} and include a
  combination of gradient ascent and simulated annealing.

Here are a few facts common to this family of optimization
schemes. First, as was proved in the case of information maximization
based on Kullback-Leibler divergence~\cite{Sharpee04}, the merit
function $F^{(\alpha)}[{\bf v}]$ does not change with the length of
the vector ${\bf v}$. Therefore ${\bf v} \cdot \nabla_{\bf v}F =0 $,
as can also be seen directly from Eq.~(\ref{gradF}), because ${\bf
v}\cdot \langle {\bf s}|x,{\rm spike}\rangle =x$ and ${\bf
v}\cdot\langle {\bf s}|x\rangle=x$. Second, the gradient is $0$ when
evaluated along the true receptive field.  This is because for the
true relevant dimension according to which spikes were generated,
$\langle {\bf s}|s_1, {\rm spike}\rangle=\langle {\bf s}|s_1\rangle $,
a consequence of the fact that relevant projections completely
determine the spike probability. 
Third, merit functions,
including variance and information, can be computed with respect to
multiple dimensions by keeping track of stimulus projections on all
the relevant dimensions when forming probability distributions
(\ref{Pv}). For example, in the case of two dimensions ${\bf v}_1$ and
${\bf v}_2$, we would use
\begin{eqnarray}
\label{multiP}
&&P_{{\bf v}_1,{\bf v}_2}(x_1,x_2|{\rm spike})=
  \int d{\bf s}\, \delta(x_1-{\bf s}\cdot {\bf v}_1)  \delta(x_2-{\bf s}\cdot {\bf v}_2)
 P({\bf    s}| {\rm spike}), \nonumber \\
&&P_{{\bf v}_1,{\bf v}_2}(x_1,x_2)=\int d{\bf s} \,
  \delta(x_1-{\bf s}\cdot {\bf v}_1)   \delta(x_2-{\bf s}\cdot {\bf v}_2)
P({\bf s}),
\end{eqnarray}
to compute the variance with respect to the two dimensions as $F[{\bf
v}_1,{\bf v}_2]=\int dx_1 dx_2 \left[P(x_1,x_2|{\rm
spike})\right]^2/P(x_1,x_2).$

If multiple stimulus dimensions are relevant for eliciting the neural
response, they can always be found (provided sufficient number of
responses have been recorded) by optimizing the variance according to
Eq.~(\ref{multiP}) with the correct number of dimensions. In practice
this involves finding a single relevant dimension first, and then
iteratively increasing the number of relevant dimensions considered
while adjusting the previously found relevant dimensions. The amount
by which relevant dimensions need to be adjusted is proportional to
the contribution of subsequent relevant dimensions to neural spiking
(the corresponding expression has the same functional form as that
for relevant dimensions found by maximizing information, cf. Appendix
B~\cite{Sharpee04}). If stimuli are either uncorrelated or correlated
but Gaussian, then the previously found dimensions do not need to be
adjusted when additional dimensions are introduced. All of the
relevant dimensions can be found one by one, by always searching only
for a single relevant dimension in the subspace orthogonal to the
relevant dimensions already found.

\section{Illustration for a model simple cell}

\begin{figure}[ht]
\begin{center}
\includegraphics[width=0.99\linewidth]{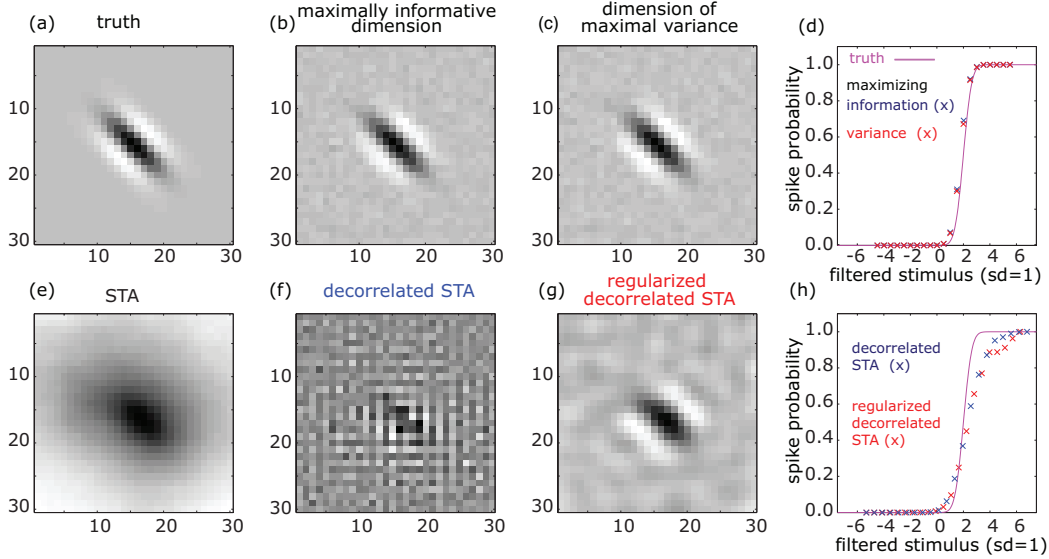}
\end{center}
\vspace*{-0.2in}
\caption{Analysis of a model visual neuron with one relevant dimension
  shown in (a). Panels (b) and (c) show normalized vectors $ \hat
  v_{\rm max}$ found by maximizing information and variance,
  respectively; (d) The probability of a spike $P({\rm spike}|{\bf s}
  \cdot \hat v_{max})$ (blue crosses -- information maximization, red
  crosses -- variance maximization) is compared to $P({\rm
  spike}|s_1)$ used in generating spikes (solid line).  Parameters of
  the model are $\sigma= 0.5$ and $\theta= 2$, both given in units of
  standard deviation of $s_1$, which is also the units for the
  $x$-axis in panels (d and h).  The spike--triggered average (STA) is
  shown in (e).  An attempt to remove correlations according to the
  reverse correlation method, $C_{a\, priori}^{-1} {\bf v}_{\rm sta}$
  (decorrelated STA), is shown in panel (f) and in panel (g) with
  regularization (see text). In panel (h), the spike probabilities as a function
  of stimulus projections onto the dimensions obtained as decorrelated
  STA (blue crosses) and regularized decorrelated STA (red crosses)
  are compared to a spike probability used to generate spikes (solid
  line). }
\label{fig:simple}
\vspace*{-0.2in}
\end{figure}

Here we illustrate how relevant dimensions can be found by maximizing
variance (equivalent to least square fitting), and compare this scheme
with that of finding relevant dimensions by maximizing information, as
well as with those that are based upon computing the spike-triggered
average.  Our goal is to reconstruct relevant dimensions of neurons
probed with inputs of arbitrary statistics. We used stimuli derived
from a natural visual environment~\cite{Sharpee06} that are known to
strongly deviate from a Gaussian distribution. All of the studies have
been carried out with respect to model neurons. Advantage of doing so
is that the relevant dimensions are known. The example model neuron is
taken to mimic properties of simple cells found in the primary visual
cortex.  It has a single relevant dimension, which we will denote as
$\hat e_1$. As can be seen in Fig.~\ref{fig:simple}(a), it is phase
and orientation sensitive.  In this model, a given stimulus ${\bf s}$
leads to a spike if the projection $s_1={\bf s} \cdot \hat e_1$
reaches a threshold value $\theta$ in the presence of noise: $P({\rm
  spike}|{\bf s})/P({\rm spike})\equiv g(s_1)=\langle H ( s_1-\theta
+\xi)\rangle$, where a Gaussian random variable $\xi$ with variance
$\sigma^2$ models additive noise, and the function $H(x)=1$ for $x>0$,
and zero otherwise.  The parameters $\theta$ for threshold and the
noise variance $\sigma^2$ determine the input--output function. In
what follows we will measure these parameters in units of the standard
deviation of stimulus projections along the relevant dimension. In
these units, the signal-to-noise ratio is given by $\sigma$.

Figure~\ref{fig:simple} shows that it is possible to obtain a good
estimate of the relevant dimension $\hat e_1$ by maximizing either
information, as shown in panel (b), or variance, as shown in panel(c).
The final value of the projection depends on the size of the dataset, as
will be discussed below.  In the example shown in
Fig.~\ref{fig:simple} there were $\approx 50,000$ spikes with average
probability of spike $\approx 0.05$ per frame, and the reconstructed
vector has a projection ${\hat v}_{max}\cdot \hat e_1= 0.98$ when
maximizing either information or variance.  Having estimated the
relevant dimension, one can proceed to sample the nonlinear
input--output function. This is done by constructing histograms for
$P({\bf s} \cdot {\hat v}_{\rm max})$ and $P({\bf s} \cdot {\hat
v}_{\rm max}|{\rm spike})$ of projections onto vector $\hat v_{\rm
max}$ found by maximizing either information or variance, and taking
their ratio. Because of the Bayes' rule, this yields the nonlinear
input--output function $g$ of Eq.~(\ref{ior}). In
Fig.~\ref{fig:simple}(d) the spike probability of the reconstructed
neuron $P({\rm spike}|{\bf s} \cdot {\hat v}_{\rm max})$ (crosses) is
compared with the probability $P({\rm spike}|s_1)$ used in the model
(solid line). A good match is obtained.

In actuality, reconstructing even just one relevant dimension from
neural responses to correlated non-Gaussian inputs, such as those
derived from real-world, is not an easy problem. This fact can be
appreciated by considering the estimates of relevant dimension
obtained from the spike-triggered average (STA) shown in panel (e).
Correcting the STA by second-order correlations of the input ensemble
through a multiplication by the inverse covariance matrix results in a
very noisy estimate, shown in panel (f). It has a projection value of
0.25.  Attempt to regularize the inverse of covariance matrix results
in a closer match to the true relevant
dimension~\cite{Theunissen00,Theunissen01,Sen01,Smyth03,Felsen05} and
has a projection value of 0.8, as shown in panel (g). While it appears
to be less noisy, the regularized decorrelated STA can have systematic
deviations from the true relevant
dimensions~\cite{Ringach97,Ringach02,Sharpee04,Sharpee06}.  Preferred
orientation is less susceptible to distortions than the preferred
spatial frequency~\cite{Felsen05}. In this case regularization was
performed by setting aside 1/4 of the data as a test dataset, and
choosing a cutoff on the eigenvalues of the input covariances matrix
that would give the maximal information value on the test
dataset~\cite{Theunissen01,Felsen05}.

\section{Comparison of Performance with Finite Data}

In the limit of infinite data the relevant dimensions can be found by
maximizing variance, information, or other objective
functions~\cite{Paninski03}.  In a real experiment, with a dataset of
finite size, the optimal vector found by any of the R\'enyi
divergences ${\hat v}$ will deviate from the true relevant dimension
$\hat e_1$. In this section we compare the robustness of optimization
strategies based on R\'enyi divergences of various orders, including
least squares fitting ($\alpha=2$) and information maximization
($\alpha=1$), as the dataset size decreases and/or neural noise
increases.

The deviation from the true relevant dimension $\delta {\bf v}={\hat
  v}-\hat e_1$ arises because the probability distributions (\ref{Pv})
are estimated from experimental histograms and differ from the
distributions found in the limit of infinite data size. The effects of
noise on the reconstruction can be characterized by taking the dot
product between the relevant dimension and the optimal vector for a
particular data sample: ${\hat v} \cdot \hat e_1=1- \frac{1}{2} \delta
{\bf v}^2$, where both ${\hat v}$ and $\hat e_1$ are normalized, and
$\delta {\bf v}$ is by definition orthogonal to $\hat e_1$.  Assuming
that the deviation $\delta v$ is small, we can use quadratic
approximation to expand the objective function (obtained with finite
data) near its maximum. This leads to an expression $\delta {\bf v} =-
[H^{(\alpha)}]^{-1} \nabla F^{(\alpha)}$, which relates deviation
$\delta v$ to the gradient and Hessian of the objective function
evaluated at the vector $\hat e_1$. Subscript $(\alpha)$ denotes the
order of the R\'enyi divergence used as an objective function.
Similarly to the case of optimizing information~\cite{Sharpee04}, the
Hessian of R\'enyi divergence of arbitrary order when evaluated along
the optimal dimension $\hat e_1$ is given by
\begin{equation}
\label{Hmtx}
 H^{(\alpha)}_{ij}=-\alpha\int dx P(x|{\rm spike})C_{ij}(x) \left[ \frac{P(x|{\rm
        spike})}{P(x)}\right]^{\alpha-3}\left [\frac{d}{dx}\left( \frac{P(x|{\rm
        spike})}{P(x)} \right) \right ]^2,
\end{equation}
where $ C_{ij}(x)=\left( \langle s_is_j|x\rangle -\langle s_i|x\rangle
  \langle s_j|x\rangle \right)$ are covariance matrices of inputs
sorted by their projection $x$ along the optimal dimension.

When averaged over possible outcomes of $N$ trials, the gradient is
zero for the optimal direction. In other words, there is no specific
direction towards which the deviations $\delta {\bf v}$ are biased.
Next, in order to measure the expected spread of optimal dimensions
around the true one $\hat e_1$, we need to evaluate $\langle \delta
{\bf v}^2\rangle={\rm Tr}\left[\langle
\nabla F^{(\alpha)} \nabla F^{(\alpha)T}\rangle
\left[H^{(\alpha)}\right]^{-2}\right] $, and therefore need to know
the variance of the gradient of $F$ averaged across different
equivalent datasets.  Assuming that the probability of generating a
spike is independent for different bins, we find that $\langle \nabla
F^{(\alpha)}_i \nabla F^{(\alpha)}_j \rangle=B^{(\alpha)}_{ij}/N_{\rm
spike}$, where
\begin{equation}
\label{Dmtx}
B^{(\alpha)}_{ij} =\alpha^2 \int dx P(x|{\rm spike}) C_{ij}(x) \left [
\frac{P(x|{\rm spike})}{P(x)} \right ]^{2\alpha -4}\left [\frac{d}{dx}
\frac{P(x|{\rm spike})}{P(x)} \right ]^2.
\end{equation} 
Therefore an expected error in the reconstruction of the optimal
filter by maximizing variance is inversely proportional to the number
of spikes:
\begin{equation}
\label{var_error}
{\hat v} \cdot \hat e_1 \approx 1-\frac{1}{2}\langle \delta {\bf
  v}^2\rangle= 1-\frac{{\rm Tr}' [B H^{-2}] }{2 N_{\rm
spike}}, 
\end{equation}
where we omitted superscripts $^{(\alpha)}$ for clarity. ${\rm Tr}'$
denotes the trace taken in the subspace orthogonal to the relevant
dimension (deviations along the relevant dimension have no
meaning~\cite{Sharpee04}, which mathematically manifests itself in
dimension $\hat e_1$ being an eigenvector of matrices $H$ and $B$ with
the zero eigenvalue). Note that when $\alpha=1$, which corresponds to
Kullback-Leibler divergence and information maximization, $A\equiv
H^{\alpha=1}=B^{\alpha=1}$. The asymptotic errors in this case are
completely determined by the trace of the Hessian of information,
$\langle \delta {\bf v}^2\rangle \propto {\rm
Tr}'\left[A^{-1}\right]$, reproducing the previously published result
for maximally informative dimensions~\cite{Sharpee04}. Qualitatively,
the expected error $\sim D/(2N_{\rm spike})$ increases in proportion
to the dimensionality $D$ of inputs and decreases as more spikes are
collected.  This dependence is in common with expected errors of
relevant dimensions found by maximizing information ~\cite{Sharpee04},
as well as methods based on computing the spike-triggered average both
for white noise~\cite{Paninski03,Rust05,Schwartz06} and correlated
Gaussian inputs~\cite{Sharpee04}.

Next we examine which of the R\'enyi divergences provides the smallest
asymptotic error~(\ref{var_error}) for estimating relevant dimensions.
Representing the covariance matrix as
$C_{ij}(x)=\gamma_{ik}(x)\gamma_{jk}(x)$ (exact expression for
matrices $\gamma$ will not be needed), we can express the Hessian
matrix $H$ and covariance matrix for the gradient $B$ as averages with
respect to probability distribution $P(x|{\rm spike})$:
\begin{equation}
B=\int dx P(x|{\rm spike})b(x)b^T(x), \quad H=\int dx P(x|{\rm
spike})a(x)b^T(x),
\end{equation}
where the gain function $g(x)=P(x|{\rm spike})/P(x)$, and matrices $
b_{ij}(x)=\alpha \gamma_{ij}(x)g'(x)\left[g(x)\right]^{\alpha-2}$ and
$a_{ij}(x)=\gamma_{ij}(x)g'(x)/g(x)$. Cauchy-Schwarz identity for
scalar quantities states that, $\langle b^2\rangle/\langle ab \rangle^2
\geq 1/\langle a^2\rangle$, where the average is taken with respect to
some probability distribution. A similar result can also be proven for
matrices under a Tr operation as in Eq. (\ref{var_error}).
Applying the matrix-version of the Cauchy-Schwarz identity to
Eq.~(\ref{var_error}), we find that the smallest error is obtained
when
\begin{equation}
{\rm Tr}'[BH^{-2}]={\rm Tr}'[A^{-1}], \quad {\rm with} \quad A=\int dx
P(x|{\rm spike})a(x)a^T(x),
\end{equation}
Matrix $A$ corresponds to the Hessian of the merit
function for $\alpha=1$: $A=H^{(\alpha=1)}$. Thus, among the various
optimization strategies based on R\'enyi divergences, Kullback-Leibler
divergence ($\alpha=1$) has the smallest asymptotic errors. The least
square fitting corresponds to optimization based on R\'enyi divergence
with $\alpha=2$, and is expected to have larger errors than
optimization based on Kullback-Leibler divergence ($\alpha=1$)
implementing information maximization. This result agrees with recent
findings that Kullback-Leibler divergence is the best distortion
measure for performing lossy compression~\cite{Tishby2007}.

\begin{figure}[ht]
\begin{center}
\includegraphics[width=0.99\linewidth]{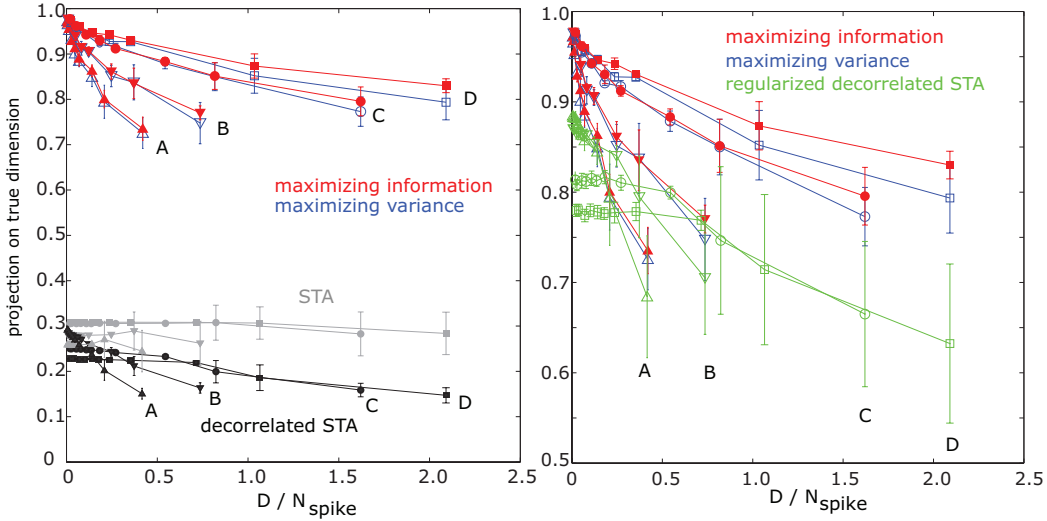}
\end{center}
\vspace*{-0.2in}
\caption{Projection of vector $\hat v_{\rm max}$ obtained by
maximizing information (red filled symbols) or variance (blue open
symbols) on the true relevant dimension $\hat e_1$ is plotted as a
function of ratio between stimulus dimensionality $D$ and the number
of spikes $N_{\rm spike}$, with $D=900$.  Simulations were carried out
for model visual neurons with one relevant dimension from
Fig.~\protect\ref{fig:simple}(a) and the input-output function
Eq.(\ref{ior}) described by threshold $\theta=2.0$ and noise standard
deviation $\sigma=1.5, \,1.0,\,0.5,\,0.25$ for groups labeled A
($\triangle$), B ($\bigtriangledown$), C ($\bigcirc$), and D ($\Box$),
respectively.  The left panel also shows results obtained using
spike-triggered average (STA, gray) and decorrelated STA (dSTA,
black).  In the right panel, we replot results for information and
variance optimization together with those for regularized decorrelated
STA (RdSTA, green open symbols). All error bars show standard
deviations.
}
\vspace*{-0.2in}
\label{fig:scaling}
\end{figure}

Below we use numerical simulations with model cells to compare the
performance of information ($\alpha=1$) and variance ($\alpha=2$)
maximization strategies in the regime of relatively small numbers of
spikes. We are interested in the range $0.1\alt D/N_{\rm spike} \alt 1
$, where the asymptotic results do not necessarily apply. The results
of simulations are shown in Fig.~\ref{fig:scaling} as a function of
$D/N_{\rm spike}$, as well as with varying neural noise levels.  To
estimate sharper (less noisy) input/output functions with
$\sigma=1.5$, $1.0$, $0.5$, $0.25$, we used larger number of bins
($16$, $21$, $32$, $64$), respectively.  Identical numerical
algorithms, including the number of bins, were used for maximizing
variance and information. The relevant dimension for each simulated
spike train was obtained as an average of 4 jackknife estimates
computed by setting aside 1/4 of the data as a test set.  Results are
shown after 1000 line optimizations ($D=900$), and performance on the
test set was checked after every line optimization.  As can be seen,
generally good reconstructions with projection values $\agt 0.7$ can
be obtained by maximizing either information or variance, even in the
severely undersampled regime $D<N_{\rm spike}$.  We find that
reconstruction errors are comparable for both information and variance
maximization strategies, and are better or equal (at very low spike
numbers) than STA-based methods. Information maximization achieves
significantly smaller errors than the least-square fitting, when we
analyze results for all simulations for four different models cells
and spike numbers ($p<10^{-4}$, paired t-test).

\section{Conclusions}

In this paper we compared accuracy of a family of optimization
strategies for analyzing neural responses to natural stimuli based on
R\'enyi divergences. Finding relevant dimensions by maximizing one of
the merit functions, R\'enyi divergence of order 2, corresponds to
fitting the linear-nonlinear model in the least-square sense to neural
spike trains. Advantage of this approach over standard least square
fitting procedure is that it does not require the nonlinear gain
function to be invertible. We derived errors expected for relevant
dimensions computed by maximizing R\'enyi divergences of arbitrary
order in the asymptotic regime of large spike numbers. The smallest
errors were achieved not in the case of (nonlinear) least square
fitting of the linear-nonlinear model to the neural spike trains
(R\'enyi divergence of order 2), but with information maximization
(based on Kullback-Leibler divergence). Numeric simulations on the
performance of both information and variance maximization strategies
showed that both algorithms performed well even when the number of
spikes is very small. With small numbers of spikes, reconstructions
based on information maximization had also slightly, but
significantly, smaller errors those of least-square fitting.  This
makes the problem of finding relevant dimensions, together with the
problem of lossy compression~\cite{Tishby,Tishby2007}, one of examples where
information-theoretic measures are no more data limited than those
derived from least squares. It remains possible, however, that other
merit functions based on non-polynomial divergence measures could
provide even smaller reconstruction errors than information
maximization.

{\small \bibliography{g:/tex/info} 

\begin{thebibliography}{10}

\bibitem{Paninski03}
L.~Paninski.
\newblock Convergence properties of three spike-triggered average techniques.
\newblock {\it Network: Comput. Neural Syst.}, 14:437--464, 2003.

\bibitem{Sharpee04}
T.~Sharpee, N.C. Rust, and W.~Bialek.
\newblock Analyzing neural responses to natural signals: Maximally informatiove
  dimensions.
\newblock {\it Neural Computation}, 16:223--250, 2004.
\newblock See also physics/0212110, and a preliminary account in {\em Advances
  in Neural Information Processing 15} edited by S. Becker, S. Thrun, and K.
  Obermayer, pp.~261-268 (MIT Press, Cambridge, 2003).

\bibitem{Tishby2007}
Peter Harremo{\"{e}}s and Naftali Tishby.
\newblock The Information bottleneck revisited or how to choose a good
  distortion measure.
\newblock {\it Proc. of the IEEE Int. Symp. on Information Theory (ISIT)},
  2007.

\bibitem{deBoer}
E.~de~Boer and P.~Kuyper.
\newblock Triggered correlation.
\newblock {\it IEEE Trans. Biomed. Eng.}, 15:169--179, 1968.

\bibitem{Hunter_Korenberg_1986b}
I.~W. Hunter and M.~J. Korenberg.
\newblock The identification of nonlinear biological systems: Wiener and
  Hammerstein cascade models.
\newblock {\it Biol. Cybern.}, 55:135--144, 1986.

\bibitem{stcov}
R.~R. de~Ruyter~van Steveninck and W.~Bialek.
\newblock Real-time performance of a movement-sensitive neuron in the blowfly
  visual system: coding and information transfer in short spike sequences.
\newblock {\it Proc. R. Soc. Lond. B}, 265:259--265, 1988.

\bibitem{Marmarelis_1997}
V.~Z. Marmarelis.
\newblock Modeling Methodology for Nonlinear Physiological Systems.
\newblock {\it Ann. Biomed. Eng.}, 25:239--251, 1997.

\bibitem{rob+bill-feature}
W.~Bialek and R.~R. de~Ruyter~van Steveninck.
\newblock Features and dimensions: Motion estimation in fly vision.
\newblock q-bio/0505003, 2005.

\bibitem{Ringach97}
D.~L. Ringach, G.~Sapiro, and R.~Shapley.
\newblock A subspace reverse-correlation technique for hte study of visual
  neurons.
\newblock {\it Vision Res.}, 37:2455--2464, 1997.

\bibitem{Ruderman}
D.~L. Ruderman and W.~Bialek.
\newblock Statistics of natural images: scaling in the woods.
\newblock {\it Phys. Rev. Lett.}, 73:814--817, 1994.

\bibitem{Sharpee06}
T.~O. Sharpee, H.~Sugihara, A.~V. Kurgansky, S.~P. Rebrik, M.~P. Stryker, and
  K.~D. Miller.
\newblock Adaptive filtering enhances information transmission in visual
  cortex.
\newblock {\it Nature}, 439:936--942, 2006.

\bibitem{Ali_Silvey_1966}
S.~M. Ali and S.~D. Silvey.
\newblock A general class of coefficeint of divergence of one distribution from
  another.
\newblock {\it J. R. Statist. Soc. B}, 28:131--142, 1966.

\bibitem{Csiszar_1967}
I.~Csisz\'ar.
\newblock Information-type measures of difference of probability distrbutions
  and indirect observations.
\newblock {\it Studia Sci. Math. Hungar.}, 2:299--318, 1967.

\bibitem{B00a}
N.~Brenner, S.~P. Strong, R.~Koberle, W.~Bialek, and R.~R. de~Ruyter~van
  Steveninck.
\newblock Synergy in a neural code.
\newblock {\it Neural Computation}, 12:1531--1552, 2000.
\newblock See also physics/9902067.

\bibitem{Theunissen00}
F.~E. Theunissen, K.~Sen, and A.~J. Doupe.
\newblock Spectral-temporal receptive fields of nonlinear auditory neurons
  obtained using natural sounds.
\newblock {\it J. Neurosci.}, 20:2315--2331, 2000.

\bibitem{Theunissen01}
F.E. Theunissen, S.V. David, N.C. Singh, A.~Hsu, W.E. Vinje, and J.L. Gallant.
\newblock Estimating spatio-temporal receptive fields of auditory and visual
  neurons from their responses to natural stimuli.
\newblock {\it Network}, 3:289--316, 2001.

\bibitem{Sen01}
K.~Sen, F.~E. Theunissen, and A.~J. Doupe.
\newblock Feature analysis of natural sounds in the songbird auditory
  forebrain.
\newblock {\it J. Neurophysiol.}, 86:1445--1458, 2001.

\bibitem{Smyth03}
D.~Smyth, B.~Willmore, G.~E. Baker, I.~D. Thompson, and D.~J. Tolhurst.
\newblock The receptive fields organization of simple cells in the primary
  visual cortex of ferrets under natural scene stimulation.
\newblock {\it J. Neurosci.}, 23:4746--4759, 2003.

\bibitem{Felsen05}
G.~Felsen, J.~Touryan, F.~Han, and Y.~Dan.
\newblock Cortical sensitivity to visual features in natural scenes.
\newblock {\it PLoS Biol.}, 3:1819--1828, 2005.

\bibitem{Ringach02}
D.~L. Ringach, M.~J. Hawken, and R.~Shapley.
\newblock Receptive field structure of neurons in monkey visual cortex revealed
  by stimulation with natural image sequences.
\newblock {\it Journal of Vision}, 2:12--24, 2002.

\bibitem{Rust05}
N.~C. Rust, O.~Schwartz, J.~A. Movshon, and E.~P. Simoncelli.
\newblock Spatiotemporal elements of macaque V1 receptive fields.
\newblock {\it Neuron}, 46:945--956, 2005.

\bibitem{Schwartz06}
Schwartz O., J.W. Pillow, N.C. Rust, and E.~P. Simoncelli.
\newblock Spike-triggered neural characterization.
\newblock {\it Journal of Vision}, 176:484--507, 2006.

\bibitem{Tishby}
N.~Tishby, F.~C. Pereira, and W.~Bialek.
\newblock The information bottleneck method.
\newblock In B.~Hajek and R.~S. Sreenivas, editors, {\it Proceedings of the
  37th Allerton Conference on Communication, Control and Computing}, pp
  368--377. University of Illinois, 1999.
\newblock See also physics/0004057.

\end{thebibliography}
\end{document}